\documentclass[11pt]{article}%%%%where rstrans is the template name
\usepackage{fullpage}
\usepackage{bm}
\usepackage[numbers]{natbib}
\usepackage{amssymb,fullpage,amsmath,amsfonts}
\usepackage{graphicx}

\newcommand{\beq}{\begin{equation}}
\newcommand{\eeq}{\end{equation}}
\newcommand{\com}{\, ,}
\newcommand{\per}{\, .}

\newcommand{\defn}{\ensuremath{\stackrel{\mathrm{def}}{=}}}

\newcommand{\dd}{\mathrm{d}}

\newcommand{\minfty}{-\infty}
\newcommand{\intfty}{\int^0_{-\infty} \!\!\!\!\!\!}
\newcommand{\atz}{\big|_0}
\newcommand{\Batz}{\Big|_0}

\newcommand{\ddxi}{(\bxio \!\bcdot \! \grad)}
\newcommand{\ddu}{(\buo \!\bcdot \! \grad)}

\newcommand{\glm}{{glm}}

\newcommand{\buS}{\bm{u}^{\mathrm{\scriptscriptstyle{S}}}}
\newcommand{\buSsol}{\bm{u}^{\mathrm{\scriptscriptstyle{S}}}_{\textrm{sol}}}
\newcommand{\buL}{\bm{u}^\mathrm{\scriptscriptstyle{L}}}

\newcommand{\buE}{\bm{u}^\mathrm{\scriptscriptstyle{E}}}

\newcommand{\uS}{u^\mathrm{\scriptscriptstyle{S}}}
\newcommand{\vS}{v^\mathrm{\scriptscriptstyle{S}}}
\newcommand{\uSsol}{u^\mathrm{\scriptscriptstyle{S}}_{\text{sol}}}
\newcommand{\vSsol}{v^\mathrm{\scriptscriptstyle{S}}_{\text{sol}}}
\newcommand{\wS}{w^\mathrm{\scriptscriptstyle{S}}}
\newcommand{\wSsol}{w^\mathrm{\scriptscriptstyle{S}}_{\text{sol}}}

\newcommand{\bTSsol}{\bm{T}^\mathrm{\scriptscriptstyle{S}}_{\text{sol}}}
\newcommand{\bTS}{\bm{T}^\mathrm{\scriptscriptstyle{S}}}

\newcommand{\nuaL}{\nu_\mathrm{a}^\mathrm{\scriptscriptstyle{L}}}
\def\CL{\mathcal{C}^\mathrm{\scriptscriptstyle{L}}}

\newcommand{\bxL}{\bm{x}^\mathrm{\scriptscriptstyle{L}}}
\newcommand{\bphiL}{\bm{\varphi}^\mathrm{\scriptscriptstyle{L}}}
\newcommand{\piL}{\pi^\mathrm{\scriptscriptstyle{L}}}
\newcommand{\pitL}{\tilde{\pi}^\mathrm{\scriptscriptstyle{L}}}

\newcommand{\ep}{\epsilon}

\newcommand{\ord}{O}

\newcommand{\bx}{\bm{x}}
\newcommand{\bxi}{\bm{\xi}}
\newcommand{\bxio}{\bm{\xi}_1}

\newcommand{\bu}{\bm{u}}
\newcommand{\buo}{\bm{u}_1}

\renewcommand{\bf}{\bm{f}}

\def\bphi{\bm{\varphi}}
\def\bXi{\bm{\Xi}}
\def\bw{\bm{w}}
\def\bq{\bm{q}}
\def\lie{\mathcal{L}}

\newcommand{\half}{\tfrac{1}{2}}

\newcommand{\p}{\partial}
\newcommand{\cross}{\times}
\newcommand{\grad}{\bm{\nabla}}

\newcommand{\curl}{\bm{\nabla} \!\times\!}
\newcommand{\bcdot}{ \bm{\cdot} }
\renewcommand{\div}{\bm{\nabla} \!\bcdot\!}
\newcommand{\lap}{\triangle}
\newcommand{\eye}{\bm{\hat x}}
\newcommand{\jay}{\bm{\hat y}}
\newcommand{\kay}{\bm{\hat z}}

\newcommand{\quarter}{\tfrac{1}{4}}

\title{Stokes drift and its discontents}

\author{%%%% Author details
Jacques Vanneste$^{1}$, William R. Young$^{2}$}
\date{\small $^{1}$School of Mathematics and Maxwell Institute for Mathematical Sciences,  University of Edinburgh, Edinburgh EH9 3FD, UK \\
$^{2}$Scripps Institution of Oceanography, University of California at San Diego, La Jolla CA 92093-0213, USA} %\\
\begin{document}

\maketitle

%%%%%%%%%% Abstract text to be placed here %%%%%%%%%%%%
\begin{abstract}
The Stokes velocity $\buS$, defined approximately by Stokes (1847,  \textit{Trans. Camb. Philos. Soc.}, \textbf{8}, 441--455),  and exactly via the Generalized Lagrangian Mean,  is divergent even in an incompressible fluid. We show that the Stokes velocity can be naturally decomposed into a solenoidal component, $\buSsol$, and a remainder that is small for waves with slowly varying amplitudes. We further show that $\buSsol$ arises as the sole Stokes velocity when the Lagrangian mean flow is suitably redefined to ensure its exact incompressibility. The construction is an application  of Soward \& Roberts's glm theory (2010, \textit{J. Fluid Mech.}, \textbf{661}, 45--72) which we specialise to surface gravity waves and  implement effectively using a Lie series expansion.  We further show that the corresponding Lagrangian-mean momentum equation is formally identical to the Craik--Leibovich equation with $\buSsol$ replacing $\buS$, and we discuss the form of the Stokes pumping associated with both $\buS$ and $\buSsol$. 

\end{abstract}
%%%%%%%%%%%%%%%% Abstract above %%%%%%%%%%%%%

\section{Introduction}
%%%% Insert A head here

Surface gravity waves induce  a rectified motion of fluid particles and thus a wave-averaged difference between the mean Eulerian velocity, $\buE$, and the mean Lagrangian velocity $\buL$  \cite{stok47,vand-brei18}
\beq
\buL = \buE+\buS\per
\label{eq1}
\eeq
Above, $\buS$ is the Stokes velocity, also known as Stokes drift.  The Stokes velocity is  fundamental to understanding wave-averaged effects,  such as  the Craik--Leibovich vortex force  and  the Stokes--Coriolis force, in the wave-averaged momentum and vorticity equations  \cite{crai-leib76,leib80,hass70,huan79,mcwil-rest99}.

The Stokes velocity  can be defined exactly at finite wave amplitude using  Generalized Lagrangian  Mean (GLM) theory \cite{andr-mcin78a,buhl14}. This exact GLM $\buS$  is  rotational  and   compressible, even if the underlying fluid motion is   irrotational and  incompressible  \cite{mcin88}. Expansion   in powers of a wave-amplitude parameter $\ep$ produces the  standard approximation \cite{phil77,vand-brei18} to the Stokes velocity
\beq
\buS = \overline{\ddxi\, \buo}\per \label{eq2}
\eeq
The overbar in  \eqref{eq2} denotes  a running time mean, or phase average. 
In \eqref{eq2}, $\buo$ is the linear (first order in $\ep$)  velocity of the  wave and the associated  displacement $\bxio$  is defined by  $\p_t \bxio=\buo$ and $\overline {\bxio}=0$. (The subscript $1$ indicates first-order fields throughout.) 
The small-amplitude approximation to $\buS$ in \eqref{eq2}  is also rotational and compressible: assuming only that $\div \bxio =0$, McIntyre \cite{mcin88} shows from \eqref{eq2} that 
\beq
\div \buS = \p_t \left(\, \half \overline {\xi_{1i} \xi_{1j}} \, \right)_{,i j}\per
\label{McI}
\eeq
The time derivative of an averaged quadratic quantity in \eqref{McI} entails the same slow-modulation assumption  that underlies the concept of group velocity and so introduces a second small parameter, $\mu$. The Eulerian mean velocity,  $\buE$ in \eqref{eq1}, is  incompressible and thus the  divergent $\buS$ in \eqref{McI}  implies a divergent Lagrangian mean velocity. 

\begin{figure}
\centering\includegraphics[width=4.0in]{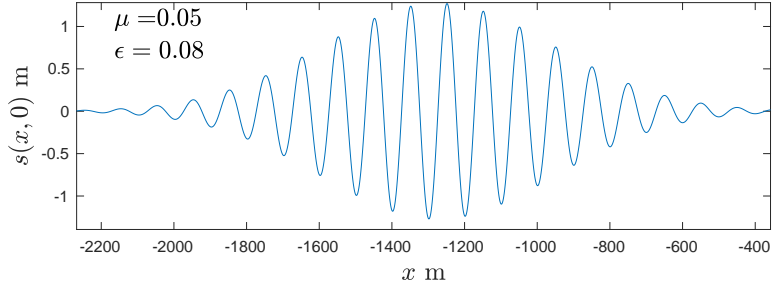}
\caption{The sea-surface displacement, $s(x,t)$, of a  packet of surface gravity waves at $t=0$.  The envelope is Gaussian, $a \exp[-(x-x_0-\half c t)^2/2 \ell^2]$, and the carrier wavenumber is $k=2 \pi/100$m. The $100$ m wave length  corresponds to an 8 second period and a group velocity $c/2$ of $6.24$ m s$^{-1}$. The maximum surface displacement $a = 1.27$ m corresponds to maximum  wave orbital speed $1$ m s$^{-1}$. The modulation parameter is $\mu =1/(k\ell)=0.05$ and the wave slope  is $\epsilon = k a =0.08$. }
\label{modFig1}
\end{figure}

\begin{figure}[!h]
% \begin{figure}
\centering\includegraphics[width=4.0in]{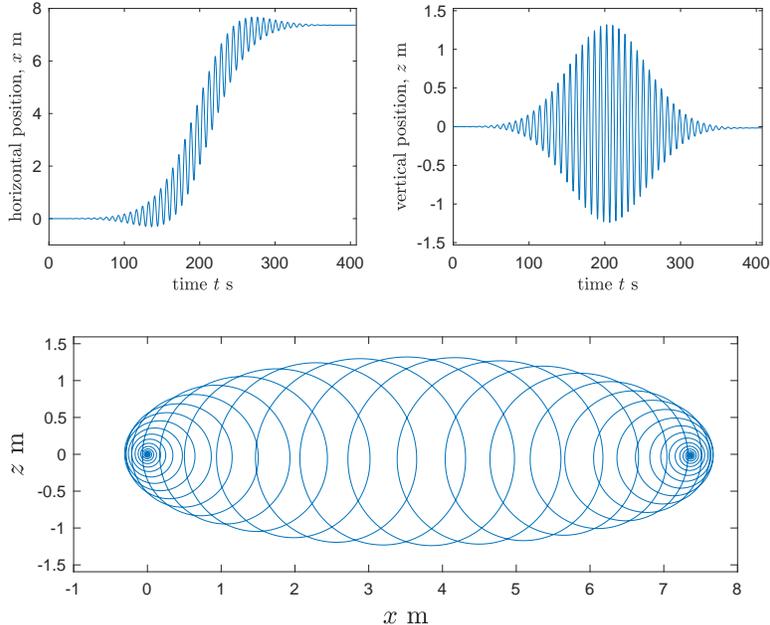}
%%% where xxxxxx name represents "figurename.eps"
\caption{Trajectory of a fluid particle in the linear  velocity field of the wave packet in figure \ref{modFig1}. The top panels show the $x$- and $z$- displacements as functions of time; the bottom panel shows the trajectory.  In this computation we assume that  the depth $d$ is much greater than the packet length scale $\ell$ so that the second-order Eulerian mean flow is negligible in the wave-active zone.}
\label{trajFig}
\end{figure}

In  figure \ref{modFig1}  we illustrate  the role of the  two small parameters $\ep$ and $\mu$, by   considering  a weakly nonlinear,  slowly modulated two-dimensional packet  of deep-water surface gravity waves. The Stokes expansion [1,10] is justified by the weakly-nonlinear assumption that the wave slope is small:  $\epsilon = ak \ll 1$, where $k$ is the wavenumber and $a$ is the amplitude of the surface displacement.  In this example  the  slow-modulation  parameter is $\mu = (k \ell)^{-1}  \ll 1$ where $\ell$ is length scale of the packet envelope. Figure \ref{trajFig} shows the motion of a fluid particle in the velocity field of this wave.

 Despite \eqref{McI}, and the exact results  provided by GLM,  some authors  are  reluctant to accept the reality of  non-zero $\div \buS$. Moreover discontent with  $\div \buS\neq 0 $ is sometimes confounded with unease over  the vertical component of the Stokes velocity, $\wS = \kay \bcdot \buS$. For example,   rather than taking the vertical component of \eqref{eq2}, McWilliams, Restrepo \& Lane \cite{mcwil-et-al}  defines a ``vertical Stokes pseudo-velocity'' which, together with the  horizontal components of $\buS$, makes an incompressible three-dimensional ``Stokes pseudo-velocity''. 
 
 Mellor \cite{mell16},  while emphasizing   that $\buS$ is divergent, is unwilling to accept  a non-zero vertical component $\wS$: ``a mean vertical drift is not acceptable''.   In view of concerns  with   the mean vertical drift $\wS$ it is reassuring  that  particle tracking velocimetry can be used to observe vertical Lagrangian displacements,  $\wS \neq 0$,  beneath groups of deep-water waves \cite{vand-et-al,moni19}:  the  mean vertical drift is upward as a wave packet arrives and downwards as the packet departs. (For the wave packet  in figure \ref{modFig1},  the maximum vertical displacement resulting from $\wS$ is about $4.8$cm -- this is not visible in figure  \ref{trajFig}.) After the passage of the packet a fluid particle returns to its initial depth. It is a \textit{net vertical displacement} that is unacceptable: transient vertical motion, on time scales longer than a ten-second wave period, and shorter than the hundred-second packet transit time,  is not a concern.

 Mellor critiques the  Craik \& Leibovich ([3,4] CL hereafter) vortex-force  formulation of wave-mean interaction by arguing that CL and subsequent authors incorrectly assume that the divergence of  $\buS$ is zero. A  different interpretation is that CL and many followers  assume that the wave field has no temporal modulation  so that the right of \eqref{McI} is conveniently zero. For example, McWilliams \& Restrepo \cite{mcwil-rest99} claims to prove $\div \buS= 0$. But examination of this argument shows that \cite{mcwil-rest99} assumes  that there is no temporal modulation of the wave field. This raises the issue of whether the CL formulation is incomplete or misleading in situations with temporal modulation of the wave field e.g. in ocean observations\cite{smith} and in modeling the growth of swell \cite{wagner}.

In this paper, we revisit the concept of Stokes velocity. For surface gravity waves we exhibit a natural Helmholtz decomposition of $\buS$ in \eqref{eq2} and argue that the solenoidal component, $\buSsol$, can advantageously replace $\buS$ in most situations. We emphasize that the familiar form \eqref{eq2} of the Stokes velocity is not unique but depends on a specific definition of the Lagrangian-mean flow, namely the GLM definition of Andrews \& McIntyre \cite{andr-mcin78a}. An alternative definition, proposed by Soward \& Roberts \cite{sowa-robe10} and closely related to classical averaging and its Lie series implementation (e.g.\ \cite{lich-lieb,nayf00}), leads to a solenoidal Lagrangian-mean velocity with $\buSsol$ as the corresponding Stokes velocity. This alternative definition, known as `\glm' but  better characterized as `solenoidal Lagrangian mean', has the added benefit of coordinate independence in any geometry, unlike standard GLM (see \cite{gilb-v18} for other coordinate-independent definitions of the Lagrangian mean). We show that, for surface gravity waves, the associated Lagrangian-mean momentum equation governing the dynamics of the Eulerian mean flow is the CL equation with $\buSsol$ replacing $\buS$.

The difference between GLM and \glm{} is vividly illustrated  with  an example proposed by O. B\"uhler (personal communication). Consider a bucket of initially motionless water. If the water is agitated, for example by pressure forcing  at the surface, then the potential energy of the water is increased, or equivalently  the center of mass of the water is   elevated  above  its initial height. The fluid at the bottom of B\"uhler's bucket, however, cannot move in the vertical and so % the vertical component of the Lagrangian motion must be divergent. Thus 
 any definition of ``Lagrangian mean'' that tracks the position of the center of mass of the fluid -- such as GLM -- will be divergent in this situation, even though the velocity of the water is entirely incompressible. Conversely, any  definition of ``Lagrangian mean''  resulting in a strictly solenoidal Lagrangian mean velocity -- such as \glm{} -- cannot track  the center of mass of the fluid.

The plan of the paper is as follows. In \S\ref{sec:Svelocity}, we sketch the derivation of the standard form \eqref{eq2} of the Stokes velocity, give its Helmholtz decomposition, and show how a simple modification of this derivation, implementing an alternative Lagrangian-mean flow definition, naturally brings about the velocity $\buSsol$. In \S\ref{Spump}, we examine the respective role of $\buS$ and $\buSsol$ in  Stokes pumping, which is the mechanism whereby the horizontal divergence of the Stokes transport drives an Eulerian mean flow. In \S\ref{sec:glm}, we show how the \glm{} approach enables the systematic construction of solenoidal Lagrangian-mean and Stokes velocities up to arbitrary algebraic accuracy in $\ep$. We explain how Lie series provide both an interpretation and an efficient implementation of this construction, and we derive the \glm{} version of the CL equations. 
Section \ref{sec:conclusion} gives conclusions.

\section{The Stokes velocity $\buS$  and its solenoidal part $\buSsol$ \label{solSec}} \label{sec:Svelocity}

\subsection{Derivation of the Stokes velocity} \label{sec:Svelocitya}

We start  by recalling the traditional derivation of the Stokes velocity in  \eqref{eq2}. The position $\bx(t)$ of a fluid particle is determined by solving
\beq
\frac{\dd \bx(t,\alpha,\ep) }{\dd t} = \bu\big(\bx(t,\alpha,\ep),t,\alpha,\ep\big)\com
\label{trad3}
\eeq
where $\alpha$ is a wave phase, regarded as %; $\alpha$ subsequently  serves as 
an ensemble parameter. 
The fluid velocity $\bu(\bx,t,\alpha,\ep)$ is incompressible, $\div \bu=0$, and  has the form 
\beq
\bu(\bx,t,\alpha,\ep) = \epsilon \bu_1(\bx,t,\alpha) +  \epsilon^2 \bu_2(\bx,t, \alpha)+ \cdots,
\label{trad7}
\eeq
where $\epsilon$ is the wave amplitude parameter. The leading order term   $\bu_1(x,t,\alpha)$ is a fast wavy flow, so $\overline{\bu_1}=0$, where the mean, denoted by the overbar, is an average over the phase $\alpha$.

To average the fast wave oscillations in \eqref{trad3} we consider  the 
\textit{ansatz}
\begin{subequations}
\begin{align}
\bx(t,\alpha,\ep) &= \bxL(t,\ep) +  \bxi \big(\bxL(t,\ep),t,\alpha,\ep\big) \com \label{trad17a} \\
&= \bxL(t,\ep) + \epsilon \bxi_1\big(\bxL(t,\ep),t,\alpha\big)+ \epsilon^2 \bxi_2 \big(\bxL(t,\ep),t,\alpha\big) + \cdots,
\label{trad17b}
\end{align}
\end{subequations}
where $\bxL(t,\ep)$ is the slow motion of a  Lagrangian mean position. Think of $\bxL$ as a `guiding center'  such  that  rapid wavy  oscillations are confined to  the displacements $\bxi_n$;  these displacements from $\bxL$ do not grow with time, i.e. all members of the ensemble remain close to the guiding center $\bxL$. The motion of $\bxL$ is written as 
\begin{align}
\frac{\dd \bxL}{\dd t} &= \ep^2 \buL  \label{trad18}\\
&= \ep^2 \buL_2(\bxL,t)+  \ep^3 \buL_3(\bxL,t) + \cdots \com
\label{trad19}
\end{align}
where  $\buL(\bx,t,\ep)$ is the Lagrangian mean velocity, yet to be defined and determined. 
The ansatz \eqref{trad17b} is ambiguous because  requiring only  that the $\bxi_n$'s do not grow with time  does not uniquely determine $\buL$ and $\bxi_n$. We return to this point below. %dwell further on this point  in section \ref{solSec}.

Substituting \eqref{trad17b} and \eqref{trad19} into \eqref{trad3} and \eqref{trad7}  and matching powers of $\epsilon$ at the first two orders results in 
\begin{subequations}
\begin{align}
\p_t \bxi_{1 } (\bxL,t,\alpha) &= \bu_{1}(\bxL,t,\alpha)\com
\label{trad23}\\
\buL_2 (\bxL,t)  + \p_t \bxi_{2}(t,\alpha) &= \bu_{2}(\bxL,t,\alpha) + \bxi_{1 }(\bxL,t,\alpha)\!\bcdot\! \grad \, \bu_{1}(\bxL,t,\alpha)\per
\label{trad29}
\end{align}
\end{subequations}
Choosing  to follow  Stokes \cite{stok47} and Andrews \& McIntyre \cite{andr-mcin78a}, one  disambiguates  \eqref{trad17b}  by requiring that $\overline{\bxi_{2}(\bxL,t,\alpha) }=0$. In this case    averaging \eqref{trad29}  produces  the familiar result
\beq
\buL = \underbrace{\overline{\bu_2} }_{\buE}+  \underbrace{\overline{ \ddxi\, \bu_{1}}}_{\buS} \qquad \text{(with $\, \overline{\bxi_2}=0$)}\per
\label{trad31}
\eeq
In \eqref{trad31}  we have now omitted the subscript 2 on $\buL$.
With \eqref{trad31} we recover  $\eqref{eq2}$ and the small wave-amplitude version of  \eqref{eq1}.

\subsection{The solenoidal Stokes velocity $\buSsol$} \label{sec:sole}

Let us examine the divergence of $\buS$ in more detail. The `un-averaged Stokes velocity' can be written exactly  as 
\begin{subequations}
\begin{align}
\bxio \bcdot \grad \buo &= \p_t \half \ddxi \bxio + \half\ddxi \buo -  \half \ddu \bxio \com
 \label{unavsd3} \\
&=  \p_t \half \ddxi \bxio  + \curl \half (\buo \cross \bxio )\per \label{eq713}
\end{align}
\end{subequations}
In passing from \eqref{unavsd3} to \eqref{eq713} we have used wave incompressibility 
%For the moment we  assume only  that the  first-order  velocity and displacement in \eqref{eq2}  are incompressible
\beq
\div \buo =\div \bxio =0\com
\label{eq5}
\eeq
to simplify  the standard vector identity for the curl of the  cross product  $ \buo \cross \bxio $.  The  average of \eqref{eq713}, 
\beq
\buS = \underbrace{\p_t  \half\overline{ \ddxi \bxio}}_{\ord(\ep^2\mu)} +  \underbrace{ \curl \half  \left( \overline{\buo \cross \bxio}\right) }_{\ord(\ep^2)}\com
 \label{sol7}
\eeq
identifies a solenoidal part of the Stokes velocity as
 \begin{subequations}
\label{uSsol}
\begin{align}
\buSsol &\defn  \curl \left( \half \, \overline{\buo \cross \bxio }\right)\com \label{7sol}\\
&= \half \overline{\ddxi \, \buo} - \half \overline{\ddu \, \bxio} \per \label{8sol}
\end{align}
\end{subequations}
The solenoidal vector $\buSsol$  is the incompressible part of the Stokes velocity for all types of weakly nonlinear waves in an incompressible fluid. We propose that $\buSsol$ can advantageously replace the traditional form of the Stokes velocity \eqref{eq2} in many circumstances.

%\medskip
%\begin{quotation}
%\noindent \textcolor{red}{A true Helmholtz decomposition
%is
%\beq
%\buS=\grad \alpha + \curl \bbeta
%\eeq
%We assume that $\buS$ is known and so are  BCs on $\buS$. Taking the  divergence 
%\begin{align}
%\lap \alpha &= \div\buS\com \\
%& = \p_t \div\left[ \quarter \grad |\bxio|^2 +\half  (\curl \bxio )\cross \bxio \right] \com \\
%&=\p_t \lap \quarter |\bxio|^2 + \p_t \div  \half \left[(\curl \bxio )\cross \bxio\right]\, .
%\end{align}
%If the final term is non-zero there is no obvious solution.
%Taking the curl
%\begin{align}
%\curl \curl \bbeta &= \curl \buS\com \\
% &= \curl \curl \half  \left( \overline{\buo \cross \bxio}\right)+ \p_t \curl \half \left[(\curl \bxio )\cross \bxio\right]\
%\end{align}
%}
%\end{quotation}

The solenoidal Stokes velocity arises naturally if a small change is made in the derivation in \S\ref{sec:Svelocity}\ref{sec:Svelocitya}: suppose   we  decide from the outset to change the definition of `Lagrangian mean' so that  $\buL$ in \eqref{trad19}  is solenoidal. To implement this choice   we disambiguate  \eqref{trad17b} with
$\overline{\bxi_2} = \half \overline{\ddxi \bxio}$ (rather than $\overline{\bxi_2}=0$). Averaging \eqref{trad29} then results in  
\beq
\buL = \underbrace{\overline{\bu_2} }_{\buE}+  \underbrace{  \curl \left( \half \, \overline{\buo \cross \bxio }\right) }_{\buSsol} \com \qquad \text{(with $ \overline{\bxi_2}=\half \overline{\ddxi  \bxio}$).}
\label{eq:modmeandispl}
\eeq
The incompressible Lagrangian mean velocity in \eqref{eq:modmeandispl} is an alternative to the traditional compressible $\buL$ in \eqref{trad31}.  The  $\ord(\ep^2)$ shift $\overline{\bxi_2}$ in the position of guiding centre $\bxL(t,\ep)$ produces   Lagrangian mean and Stokes velocities that are  divergence free at order $\epsilon^2$. In \S\ref{sec:glm} 
we discuss a systematic framework -- Soward \& Robert's\cite{sowa-robe10} glm alternative to GLM  -- that generalizes this property to arbitrarily high order in $\ep$. 

Note that the difference between $\buS$ and $\buSsol$ is a time derivative so has no impact on particle dispersion for waves that are represented by stationary random processes as discussed by Holmes-Cerfon and B\"uhler \cite{buhler2009particle}.

\subsection{Irrotational linear waves \label{irrot}}

In addition to incompressibility in \eqref{eq5}, surface gravity waves are  also  irrotational,
\beq
\curl \buo =\curl \bxio=0\per
\label{eq6}
\eeq
Using  \eqref{eq6}, the final term in the vector identity  $\bxio\bcdot \grad \bxio = \grad \half |\bxio|^2+ (\curl \bxio) \cross \bxio$ is zero and  the average of  \eqref{sol7} simplifies further to a  Helmholtz decomposition of the surface-wave Stokes velocity
\begin{align}
\buS &= \p_t \grad  \tfrac{1}{4}\overline{ |\bxio |^2}   +\underbrace{ \curl \half \overline{\buo \cross \bxio }}_{\buSsol}\per
\label{eq17}
\end{align}
The divergence of the surface-wave Stokes velocity in \eqref{eq17}  is 
\begin{align}
\div \buS &= \underbrace{\p_t \lap \,  \tfrac{1}{4} \overline{|\bxio |^2}}_{\ord(\epsilon^2 \mu)}\com \label{eq23}
\end{align}
 where $\lap = \p_x^2+\p_y^2 + \p_z^2$ is the Laplacian. The expression for $\div \buS$ in \eqref{eq23} is simpler than that in \eqref{McI}, but restricted to waves with $\curl \bxio=0$. 
 
The curl of the  Stokes  velocities does not vanish:
 \begin{align}
 \curl \buS =\curl \buSsol &= \curl \curl \half \overline{\buo \cross \bxio}\com \\
 &= - \lap \half \overline{\buo \cross \bxio}\com \label{Svort5}
 \end{align}
 where we have used $\curl \curl = \grad \div - \lap$ and $\div \half \overline{\buo \cross \bxio} =0$. With an error of order $\mu^2 \epsilon^2$ we can replace $\lap$ with $\p_z^2$ in \eqref{Svort5}.
 
% replace "The vorticity of the Stokes flow is" with "The curl of the (solenoidal) Stokes drift velocity does not vanish:". 

 \section{Stokes transport and Stokes pumping \label{Spump}}
 
Discussions of the deep return flow associated with a surface-gravity wave packet   argue \cite{vand-brei18} that: 
\begin{quotation} \noindent `` the  return flow $\cdots$ can be explained as the irrotational response to balance the Stokes transport $\cdots$ that acts to `pump' fluid from the trailing edge of the packet  to the leading edge''. \end{quotation}
Similar sentiments are expressed in \cite{higg-et-al,haney}. With  this physical picture  in mind, it is instructive to compare the Stokes pumping associated with  $\buS$ with that of $\buSsol$.  The interpretation of the return flow in the quotation  above is most consistent with $\buSsol$.

 \subsection{Stokes pumping and   $\buSsol = \curl \half \overline{\buo \cross \bxio}$}
 
 We start with the easy solenoidal case, with Stokes transport 
 \beq
\bTSsol(x,y,t) \defn \int_{\minfty}^0 \!\!\! \!\!\!   \uSsol \, \eye + \vSsol  \,  \jay   \, \dd z\com 
\eeq
where the  vertical integration above is from  the bottom of the wave-active zone (denoted $-\infty$) to the mean sea surface at $z=0$. (In this section we confine attention to deep-water waves so that the lower limit, $-\infty$, is well above the distant bottom.) Vertical integration of $\div \buSsol=0$ over the wave-active region  produces the unsurprising result 
\begin{subequations}
\begin{align}
\text{solenoidal Stokes pumping} &\defn - \div \bTSsol \com \\
& = \wSsol\atz\com \label{solPump7}
\end{align}
\end{subequations}
where we use  $\atz$ to denote evaluation at $z=0$ e.g. $ \wSsol\atz =\wSsol(x,y,0,t)$.
The result  \eqref{solPump7}  is consistent with the idea that horizontal convergence within the wave active zone pumps  fluid downwards, out of the wave-active zone, with the vertical velocity $\wSsol(x,y,0,t)$. 

Vertical integration of $\buSsol = \half \overline{(\bxio\bcdot \grad)\buo} - \half \overline{(\buo\bcdot\grad )\bxio}$ over the wave active zone  results in
\begin{subequations}
\begin{align}
\eye \bcdot \bTSsol & = \half \big( \overline{\zeta_1 u_1} - \overline{w_1 \xi_1}  \big)\atz + \p_y \chi\com \label{solPump12}\\
\jay \bcdot \bTSsol & = \half \big( \overline{\zeta_1 v_1} - \overline{w_1 \eta_1}  \big)\atz - \p_x \chi  \com
\label{solPump13}
\end{align}
\label{solPumpz}
\end{subequations}
where
\beq
\chi \defn \half  \intfty \left( \overline{\eta_1u_1} - \overline{v_1 \xi_1} \right) \, \dd z\com 
\eeq
is a streamfunction for horizontally circulating  Stokes transport. This horizontal Stokes circulation is previously unremarked, perhaps because the  terms involving $\chi$ in \eqref{solPump12} and \eqref{solPump13} are order $\mu$ smaller than the other terms. Using the coordinate expressions in  \eqref{solPump12} and \eqref{solPump13} the solenoidal  pumping can be expressed entirely in terms of surface quantities:
\beq
\text{solenoidal Stokes pumping}  = \p_x  \half \big( \overline{\zeta_1 u_1} - \overline{w_1 \xi_1}  \big)\atz+ \p_y \half \big( \overline{\zeta_1 v_1} - \overline{w_1 \eta_1}  \big)\atz\per
\label{solPump17} 
\eeq

\subsection{Stokes pumping and   $\buS = \overline{(\bxio \bcdot \grad) \buo}$}

We turn now to the traditional definition of the Stokes velocity. Define the Stokes transport via
\beq
\bTS(x,y,t) \defn \intfty \uS \, \eye + \vS \,  \jay  \, \dd z\com 
\eeq
and the Stokes pumping by
\beq
\text{Stokes pumping} \defn - \div \bTS\per
\eeq

Because $\div \buS \neq 0$ there is no analogy to \eqref{solPump7}. Instead, integrating \eqref{eq23} over the wave-active zone
\beq
\div \bTS + \wS\atz= \p_t \p_z \quarter \overline{|\bxio|^2}\atz + \p_t (\p_{x}^2 + \p_y^2) \intfty \quarter |\bxio|^2 \, \dd z\per
\label{StPump2}
\eeq  
With vertical integration of the horizontal components of $\buS$ over the wave-active zone
\begin{subequations}
\begin{align}
\eye\bcdot \bTS &=  \big(\overline{\zeta_1 u_1}\big)\atz +\p_t\p_x \intfty \half \overline{\xi_1^2} \, \dd z + \p_y \intfty \overline{\eta_1 u_1} \, \dd z \com \label{StPump23} \\
\jay \bcdot \bTS &=  \big(\overline{ \zeta_1 v_1}\big)\atz +  \p_x \intfty \overline{\xi_1 v_1} \, \dd z + \p_t \p_y \intfty \half \overline{\eta_1^2}  \dd z \per \label{StPump29}
\end{align}
\end{subequations}
The horizontal divergence of $\bTS$ is therefore
\beq
\div \bTS = \p_{x}\big(\overline{\zeta_1u_1}\big)\atz+\p_y \big(\overline{\zeta_1 v_1}\big)\atz + \p_t \Bigg(\half \intfty \bxi_{1,\alpha}  \bxi_{1,\beta}\, \dd z\Bigg)_{,\alpha \beta}\per
\label{StPump3}
\eeq
 In the final term the indices $\alpha$ and $\beta$ are $1$ and $2$.  Eliminating $\div \bTS$ between  \eqref{StPump2} and \eqref{StPump3}
 \begin{align}
 \wS\atz &= -  \p_{x}\big(\overline{\zeta_1u_1}\big)\atz- \p_y \big(\overline{\zeta_1 v_1}\big)\atz + \p_t \p_z \quarter \overline{|\bxio|^2}\atz \nonumber \\
  &  \qquad + \p_t (\p_x^2 + \p_y^2) \intfty \quarter |\bxio|^2 \, \dd z-  \p_t \Bigg(\half \intfty \bxi_{1,\alpha}  \bxi_{1,\beta}\, \dd z\Bigg)_{,\alpha \beta}\per
  \label{StPump5}
 \end{align}
 The results in \eqref{StPump2},  \eqref{StPump3} and \eqref{StPump5} are all more complicated than their solenoidal cousins in \eqref{solPump7}, \eqref{solPump12} and \eqref{solPump13}.
 
 \subsection{A comment on approximations to the Stokes transport}
  A widely used  expression for the Stokes transport is 
  \beq
   \bTS \approx \overline{(\zeta_1\buo)} \Batz\per
   \label{SapproxComment}
   \eeq
Expression \eqref{SapproxComment} is exact for a uniform ($\mu=0$) progressive wave and is a leading-order approximation for a slowly modulated ($\mu \ll 1$) wave packet. To see the connection with the more general and exact results above, align the $x$-axis with the horizontal wavevector  so that $\eta_1=v_1=0$;  then $\jay \bcdot \bTS=\jay \bcdot \bTSsol=\chi =0$. Thus, with an error by order $\epsilon^2 \mu$,   and in agreement with \eqref{SapproxComment},
\beq
\half \big( \overline{\zeta_1 u_1} - \overline{w_1 \xi_1} \big)\Batz \approx  \overline{(\zeta_1u_1)} \Batz \per
\eeq
In other words,  \eqref{SapproxComment} is a leading-order approximation  and at this order  $\bTS$ and $\bTSsol$ are identical. 
%\WRY{But approximate and special results such as \eqref{SapproxComment} cannot be fundamental. Thus we recommend \eqref{solPump12} and \eqref{solPump13}: these exact results, based on $\buSsol$, are also  simpler than the alternatives in \eqref{StPump23} and \eqref{StPump29}. }

% The final term in \eqref{StPump3} is smaller by order $\mu^2$ than the others and therefore
%\beq
% \text{Stokes pumping}\defn- \div \bTS  \approx -\p_{x}\big(\overline{\zeta_1 u_1}\big)\atz - \p_y \big(\overline{\zeta_1 v_1}\big)\atz \per
% \eeq

\section{glm} \label{sec:glm}

We return to the definition of the Stokes velocity  and show how the glm theory of Soward \& Roberts \cite{sowa-robe10} rationalizes and generalizes the heuristic construction of the solenoidal velocity $\buSsol$. % in \S\ref{sec:Svelocity}\ref{sec:sole}.
In \S\ref{sec:Svelocity}\ref{sec:Svelocitya} we emphasized the ambiguity in the decomposition \eqref{trad17a} of trajectories into a mean part $\bxL$ and a perturbation $\bxi$. GLM resolves this ambiguity by imposing that
\beq
\overline {\bxi}=0 \per
\label{eq:zeromeandispl}
\eeq
While \eqref{eq:zeromeandispl} is widely accepted, it is neither inevitable nor particularly natural. It implies that the Lagrangian-mean trajectory of a particle is defined by the equality $\bxL = \overline{\bx}$ between the coordinates of the Lagrangian-mean position and the average of the coordinates of the particle. This indicates that the Lagrangian-mean trajectory and, as a result, the Lagrangian-mean velocity $\buL$  depend on a choice of coordinates. This undesirable feature of GLM is remedied by glm,  while also ensuring that $\buL$ is non-divergent (see \cite{gilb-v18} for other alternatives to GLM).

\subsection{Formulation}

To introduce glm, it is convenient to rewrite equation \eqref{trad3} governing the  fluid trajectories in terms of
the flow map $\bphi(\bx,t,\alpha,\ep)$ giving the position at time $t$ of the fluid particle initially at $\bx$. Equation \eqref{trad3} becomes
\beq
\dot{\bphi}(\bx,t,\alpha,\ep)  = \bu(\bphi(\bx,t,\alpha,\ep),t,\alpha,\ep)\per
\label{ODE}
\eeq
The decomposition of  trajectories into mean and perturbation is best written as the composition
\beq
\bphi = \bXi  \circ \bphiL \com
\label{flowdecom}
\eeq
or, more explicitly, $\bphi(\bx,t,\alpha,\ep) = \bXi(\bphiL(\bx,t,\ep),t,\alpha,\ep)$.
Here $\bphiL$ is the ($\alpha$-independent) mean map, sending the initial position of particles to their mean position, and 
$\bXi$ is the perturbation map, sending the mean position to the exact, perturbed position. The perturbation map can only be represented in the familiar form $\bx \mapsto \bx + \bxi(\bx,t,\alpha,\ep)$, as in \eqref{trad17a}, in Euclidean space, where positions $\bx$ can be identified with vectors and added, or, in more general geometries,
once a specific coordinate system has been chosen and $\bx$ is interpreted as a triple of coordinates. The smallness of the perturbation, usually stated at $|\bxi | \ll 1$,  translates into the requirement that $\bXi$ is close to the identity map.
We emphasise that the mean map ${\bphiL}$ is not obtained from $\bphi$ by applying an averaging operator: averaging is a linear operation that applies to linear objects, such as vector fields, but not to nonlinear maps such as $\bphi$. 
Instead, $\bphiL$ is defined by imposing a condition on the perturbation map $\bXi$. The form of the Lagrangian-mean velocity $\buL$,  defined
by 
\beq
\dot{\bphiL}(\bx,t,\ep) = \ep^2 \buL\big(\bphiL(\bx,t),t,\ep\big) \com \quad \textrm{i.e.} \quad \ep^2 \buL = \dot {\bphiL} \circ (\bphiL)^{-1} \com
\label{eq:buL}
\eeq
depends on this condition. 

The defining condition of glm is expressed as follows. 
 The small parameter $\ep$ is regarded as a fictitious time, and the perturbation map $\bXi$ is constructed as the flow at `time' $\ep$
of a vector field, $\bm{q}$ say; that is, $\bXi(\bx,t,\ep)$ is the solution of 
\beq
\partial_\ep  \bXi(\bx,t,\alpha,\ep) = \bm{q}(\bXi(\bx,t,\alpha,\ep),t,\alpha,\ep) \quad \textrm{with} \ \ \bXi(\bx,t,\alpha,\ep=0) = \bx\com 
\label{eq:Xiq}
\eeq
and $t$ is treated as a fixed parameter. The glm condition is then  
\beq
\overline{\bm{q}} = 0.
\label{eq:meanq}
\eeq
Although superficially similar to the GLM condition \eqref{eq:zeromeandispl}, \eqref{eq:meanq} is fundamentally different in that it is an intrinsic statement, applicable to any manifold and independent of any coordinate choice. Moreover, the glm formulation defines an exactly divergence-free Lagrangian mean flow, by requiring that 
\beq
\grad \bcdot \bm{q} =0
\eeq
to ensure that $\bXi$ and $\bphiL$ preserve volume and hence $\div \buL = 0$.  %Note that the definition of $\bm{q}$ by \eqref{eq:Xiq} is purely formal 
%because not all maps can be written as time-$1$ flows of autonomous vector fields. This is not a major difficulty, however, since glm is implemented in a perturbative manner using that $\ep \ll 1$.  Eq.\ \eqref{eq:Xiq} 
%and defines $\bm{q}$ as an asymptotic rather than convergent series. 

Eq.\ \eqref{eq:meanq} generalizes the condition  on $\overline{\bxi_2}$ in \eqref{eq:modmeandispl} which leads to $\buSsol$  as the Stokes velocity.
We show this  by solving \eqref{eq:Xiq} by Taylor expansion. In coordinates, we have that
\beq
\bXi(\bx,t,\alpha,\ep) = \bx + \bxi(\bx,t,\alpha,\ep) = \bx + \ep \bm{q}(\bx,t,\alpha,0) + \tfrac{1}{2} \ep^2 (\partial_\ep \bq +\bm{q} \bcdot \grad \bm{q})(\bx,t,\alpha,0) + \cdots.
\eeq
The power series expansions \eqref{trad17b} for $\bxi$ and
\beq
\bq = \bq_1 + \ep \bq_2 + \ep^2 \bq_3 + \cdots %\quad \textrm{and} \quad \buL = \ep^2 \buL_2 + \cdots
\label{eq:qpower}
\eeq 
for  $\bq$ then give 
\beq 
\bxio= \bm{q}_1 \quad \textrm{and} \quad \bxi_2 = \bm{q}_2 + \tfrac{1}{2} \bm{q}_1 \bcdot \grad \bm{q}_1
\eeq
so that \eqref{eq:meanq} implies \eqref{eq:modmeandispl}. 
%This in turn defines a solenoidal Stokes velocity via $\buL = \buE + \buSsol$, with the leading-order $\buSsol$ given in \eqref{sol}.
In the next section, we develop a systematic computation of $\buL$ and hence $\buSsol$ order by order in $\ep$ using Lie series. This removes the need to introduce $\bxi$ by focussing the perturbation expansion on $\bq$.

\subsection{Lie series expansion}

The glm formalism can be regarded as an instance of classical perturbation theory, which approximates solutions to the ordinary differential equation  \eqref{ODE} by performing a variable transformation designed to eliminate fast time dependences. In this interpretation, $\bXi$ determines the variable transformation and $\bphiL$ represents the new variable. Lie series \cite{lich-lieb,nayf00} provide a powerful tool for the systematic implementation of classical perturbation theory which we now apply to  glm.  

Introducing the  decomposition \eqref{flowdecom} into \eqref{ODE} and using \eqref{eq:buL} gives
\beq
\bw + \bXi_{*} \buL = \bu,
\label{ubaru}
\eeq
where 
\beq
\bw = \dot \bXi \circ \bXi^{-1}
\label{w}
\eeq 
is the perturbation velocity and $\bXi_{*}$ is the push-forward by $\bXi$, with $\bXi_* \buL = (\buL \bcdot \grad) \bXi$ in Cartesian coordinates. Pulling back \eqref{ubaru} gives
\beq
 \buL = \bXi^* \bu - \hat  \bw, \quad \textrm{where} \ \ \hat  \bw = \bXi^* \bw. 
 \label{baru}
\eeq
We seek an $\ep$-dependent $\bXi$ to eliminate fast time dependence from $\buL$ order-by-order in $\ep$, and formulate the problem in terms of the vector field $\bm{q}$ that generates $\bXi$ according to \eqref{eq:Xiq}. 
%It is convenient to directly use $\ep$ in place of the fictitious time $s$, replacing 
%\eqref{eq:Xiq} by
%\beq
%\partial_\ep  \bXi(\bx,t,\ep) = \bm{q}(\bXi(\bx,t,\ep),t,\ep) \quad \textrm{with} \ \ \bXi(\bx,t,\ep=0) = \bm{x}.
%\label{eq:Xiq}
%\eeq
We impose that $\grad \bcdot \bq  = 0$ to ensure that $\grad \bcdot \buL = 0$, and  the glm condition \eqref{eq:meanq}. 

Expanding $\bm{q}$ as in \eqref{eq:qpower}
we relate the various terms by differentiating \eqref{baru} repeatedly with respect to $\ep$ and evaluating the results at $\ep=0$. Two key identities turns this into a mechanical exercise. 
The first, essentially the definition of the Lie derivative \cite{fran11},  is
\beq
\partial_\ep \bXi^* \bu = \bXi^* \lie_{\bq} \bu,
\label{lie}
\eeq
where
$\lie_{\bq} \bu = \bq \bcdot \grad \bu - \bu \bcdot \grad \bq$ is the Lie derivative of $\bu$ along $\bq$. The second,
\beq
\partial_\ep \hat \bw = \partial_t ( \bXi^* \bq) + \lie_{\bXi^* \bq} \hat \bw,
\label{dw}
\eeq
is established in Appendix \ref{app:details}.
Iterating \eqref{lie} and \eqref{dw} we find
\begin{align}
\bXi^* \bu &= \bu + \ep \lie_{\bq_1} \bu + \tfrac{1}{2} \ep^2 \left( \lie^2_{\bq_1} + \lie_{\bq_2} \right) \bu + \tfrac{1}{6} \ep^3 \left(\lie^3_{\bq_1} + 2 \lie_{\bq_1} \lie_{\bq_2} + \lie_{\bq_2} \lie_{\bq_1} + 2 \lie_{\bq_3} \right)  \bu + \cdots \label{xi*} \\
%&+ \tfrac{1}{24} \ep^4 \left( \lie^4_{\bq_1} + \lie_{\bq_2} \lie^2_{\bq_1} + 2 \lie_{\bq_1} \lie_{\bq_2} \lie_{\bq_1} + 3 \lie^2_{\bq_1} \lie_{\bq_2} \right. \nonumber \\
%&\qquad \qquad \left. + 2 \lie_{\bq_3} \lie_{\bq_1} + 3 \lie^2_{\bq_2} + 6 \lie_{\bq_1} \lie_{\bq_3} + 6 \lie_{\bq_4} \right) \bu + \cdots \\
\hat \bw &= \ep \partial_t \bq_1 + \tfrac{1}{2} \ep^2 \left(\partial_t \bq_2 + \lie_{\bq_1} \partial_t \bq_1 \right) 
+ \tfrac{1}{3} \ep^3 \left(  \partial_t \bq_3 + \lie_{\bq_1} \partial_t \bq_2 + \tfrac{1}{2} \lie_{\bq_2} \partial_t \bq_1 + \tfrac{1}{2} \lie^2_{\bq_1} \partial_t \bq_1 \right) + \cdots
%&+ \tfrac{1}{24} \ep^4 \left( \partial_t (\cdots) + 3 \lie_{\lie_{\bq_1} \bq_2} \hat \bw_1 + 6 \lie_{\bq_3} \hat \bw_1 + 6 \lie_{\bq_2} \hat \bw_2 + 6 \lie_{\bq_1} \hat \bw_3  \right) +   \cdots.
\label{hatw}
\end{align}
%In \eqref{hatw}, $\hat \bw_1$,   $\hat \bw_2$ and $\hat \bw_3$ on the second line should be considered as shorthand for the terms of corresponding order in the first line, so that $\hat \bw$ is expressed entirely in terms of $\bq$; we do not detail the terms in the time derivative since they do not affect $\buL$ to the order we will consider. 

Introducing \eqref{xi*}--\eqref{hatw} into \eqref{baru} gives at the first two orders in $\ep$,
\beq
\partial_t \bq_1 = \bu_1, \quad \textrm{i.e.} \quad \bq_1 = \bxi_1
\eeq
and
\beq
\buL =  \overline{\bu}_2 + \tfrac{1}{2}  \overline{\lie_{\bq_1} \bu_1} 
\eeq
on using \eqref{eq:meanq}. This provides the geometric formula
\beq
\buSsol = \tfrac{1}{2}  \overline{\lie_{\bq_1} \bu_1} 
\label{eq:buSsolgeo}
\eeq
for the solenoidal Stokes drift, equivalent to \eqref{7sol} since $\lie_{\bq_1} \bu_1 = \bq_1 \bcdot \grad \bu_1 - \bu_1 \bcdot \grad \bq_1 = \grad \times (\bu_1 \times \bxi_1)$. The expansion can be pursued to higher orders by choosing divergence-free $\bq_n$ that push the fast dependence on the right-hand side of \eqref{baru} to order $\ep^{n+1}$. This yields $\buL$ and hence $\buSsol$ to arbitrary order in $\ep$.

\subsection{Lagrangian-mean momentum equation}

A Lagrangian-mean momentum equation, analogous to Andrews \& McIntyre Theorem I \cite{andr-mcin78a}, can be derived for the glm formalism order-by-order in $\ep$ \cite{sowa-robe10,gilb-v18}. We sketch a derivation, focussing on the case of small-amplitude surface gravity waves. 
This derivation is conveniently carried out using a representation of the rotating Euler equation
\beq
D_t \bu+ \bm{f} \times \bu = - \grad p,
\label{Euler}
\eeq
with $D_t = \partial_t + \bu \bcdot \grad$, in terms of the absolute momentum $1$-form \cite{gilb-v18,holm19,holm21}
\beq
\nu_a = \bu \bcdot \dd \bx + \tfrac{1}{2} (\bm{f} \times \bx) \bcdot  \dd \bx.
\label{nuapi}
\eeq
It can be checked that \eqref{Euler} is equivalent to
\beq
(\partial_t + \lie_{\bu}) \nu_a = - \dd \pi,
\label{Eulerform}
\eeq
where $\pi = p - \tfrac{1}{2} |\bu|^2 - \tfrac{1}{2}(\bm{f} \times \bx) \bcdot \bu$, using basic properties of the Lie derivative, namely 
Leibniz rule, commutation with the differential $\dd$, and that $\lie_{\bu}= \bu \bcdot \nabla$ when applied to scalars \cite{fran11}; see Appendix \ref{app:details} for details. Equation \eqref{Eulerform} can be thought of as a local version of Kelvin's circulation theorem. This is readily obtained by integration along a closed curve $\mathcal{C}(t)$ moving with $\bu$ to find
\beq
\oint_{\mathcal{C}(t)} \nu_a = \oint_{\mathcal{C}(t)} \left( \bu  + \tfrac{1}{2} (\bm{f} \times \bx) \right) \bcdot  \dd \bx = \mathrm{const}.
\eeq
An advantage of \eqref{Eulerform} is that it directly leads to a Lagrangian-mean momentum equation of a similar form \cite{gilb-v18},
\beq
(\partial_t + \lie_{\buL}) \nuaL = - \dd \piL \com
\label{avEuler}
\eeq
and to the corresponding Lagrangian-mean Kelvin's circulation theorem
\beq
\oint_{\CL(t)} \nuaL = \mathrm{const},
\eeq
where the closed curve $\CL(t)$ moves with the Lagrangian-mean velocity $\buL$.
Here the Lagrangian-mean momentum and effective pressure are given by 
\beq
\nuaL= \overline{\bXi^* \nu_a} \quad  \textrm{and} \quad \piL = \overline{\bXi^* \pi}\per
\eeq
The pull-back $\bXi^*$ by the perturbation map acts on scalars as a composition, e.g.\ $(\bXi^* \pi)(\bx,t,\alpha,\ep)=\pi(\bXi(\bx,t,\alpha,\ep),t,\alpha,\ep)$, and commutes with the differential so that
\beq
\left(\bXi^* (\bu \bcdot \dd \bx)\right)(\bx) = \bu(\bXi(\bx)) \bcdot \dd \bXi(\bx) = \bu(\bXi(\bx)) \bcdot  (\dd \bx  \bcdot \grad) \bXi(\bx) .
\eeq

We show in Appendix \ref{app:details} that \eqref{avEuler}, together with the  coordinate representation $\bXi(\bx,t,\alpha,\ep)=\bx + \bxi(\bx,t,\alpha,\ep)$ and GLM condition \eqref{eq:zeromeandispl} recovers Andrews \& McIntyre's Lagrangian-mean momentum equation \cite[Theorem I]{andr-mcin78a}. 
%and, for small-amplitude surface waves, the Craik--Leibovich equation \eqref{wav3} (see supplementary material). 
For glm, we can use the Lie-series expression \eqref{xi*} to obtain an expansion of $\nuaL$ as
\beq
\nuaL = \tfrac{1}{2} (\bm{f} \times \bx) \bcdot \dd \bx +\ep^2 \, \overline{ \bu_2 \bcdot \dd \bx + 
 \lie_{\bq_1} (\bu_1 \bcdot \dd \bx) + \tfrac{1}{4} \lie_{\bq_1}^2 \left( (\bm{f} \times \bx) \bcdot  \dd \bx \right)} + O(\ep^3)
 \label{glmmom}
\eeq
using that $\bu = \ep \bu_1 + \ep^2 \bu_2 + \cdots$ and $\overline{\bq_1}=\overline{\bq_2} = 0$. Introducing \eqref{glmmom} into \eqref{avEuler} and expanding the Lie derivative gives an evolution equation for the Eulerian mean velocity $\buE = \overline{\bu_2}$ supplemented by the incompressibility condition $\div \buE = 0$. 

The Lagrangian-mean momentum equation greatly simplifies for surface gravity waves:  typical wave frequencies $\sigma$ satisfy $\sigma/f \gg 1$ and we can therefore neglect the term involving $\lie_{\bq_1}^2$ in \eqref{glmmom} against the other two in the average (recall that $\partial_t \bq_1 = \bu_1$). Moreover, using the irrotationality condition \eqref{eq6}, we compute
\begin{align}
\overline{\lie_{\bq_1} (\bu_1 \bcdot \dd \bx)} &= \overline{(\bq_1 \bcdot \grad ) \bu_1 \bcdot \dd \bx + \bu_1 \bcdot \dd \bq_1} = \overline{(q_{1j} u_{1i,j}  + u_{1j} q_{1j,i})} \dd x_i \nonumber \\
&= \overline{(q_{1j} u_{1j,i}  + u_{1j} q_{1j,i})} \dd x_i = \partial_t \overline{(q_{1j} q_{1j,i})} \dd x_i \nonumber
\\ &= \tfrac{1}{2} \partial_t \grad \overline{|\bq_1|^2} \cdot \dd \bx = \tfrac{1}{2} \dd (\partial_t \overline{|\bq_1|^2}) =  O(\mu \ep^2). 
\label{midterm}
\end{align}
Therefore, to leading order, the glm {mean} momentum equation reduces to
\beq
(\partial_t + \lie_{\buL}) \left( \buE \bcdot \dd \bx + \tfrac{1}{2} (\bm{f} \times \bx)   \bcdot \dd \bx \right) = - \dd \piL,
\label{aa}
\eeq
with the glm Lagrangian-mean velocity in \eqref{eq:modmeandispl} (see \cite{holm19,holm21} for an analogous formulation of the GLM CL equation). Using Cartan's formula in the form 
\beq
\lie_{\bu} ( \bm{v} \bcdot \dd \bx) = \left(( \curl \bm{v}) \times \bu \right) \bcdot \dd \bx + \dd ( \bu \cdot \bm{v}) \com 
\eeq
 the Lie derivatives in \eqref{aa} can be written as
\begin{align}
\lie_{\buL}  \left((\bm{f} \times \bx) \bcdot \dd \bx \right) &= 2 \left( \bm{f} \times \buL \right) \bcdot \dd \bx + \dd ( \cdot), \\
\lie_{\buL} \left( \buE   \bcdot \dd \bx \right) &= \lie_{\bu^\mathrm{E}} \left( \buE  \bcdot \dd \bx \right) + \lie_{\buSsol} \left( \buE  \bcdot \dd \bx \right) \nonumber \\
&= (\buE \bcdot \grad \buE) \bcdot \dd \bx + \left((\curl \buE) \times \buSsol\right) \bcdot \dd \bx  + \dd (\cdot),
\end{align}
where  we do not detail the exact differentials. This makes it possible to rewrite \eqref{aa} as 
\beq
\p_t \buE + \left(\buE\!\bcdot \!\grad\right)\,  \buE + \bf \cross (\buE+\buSsol) = -\grad\varpi + \buSsol\cross (\curl \buE)\com \label{wav3}
\eeq
for a suitable definition of the effective pressure {$\varpi$}. Equation \eqref{wav3} can be recognized as the CL equation \cite{crai-leib76,leib80,holm96} with the solenoidal Stokes velocity $\buSsol$ replacing $\buS$. This is not surprising since the $O(\ep^2 \mu)$ difference between $\buSsol$ and $\buS$ is of the same order as terms neglected in the derivation of the CL equation. (CL geared $\mu$ to $\ep$  by taking  $\mu = \ep^2$. In our derivation this gearing is not necessary.) In fact the assumption $\mu \ll 1$ is only used to neglect the term \eqref{midterm} from the Lagrangian-mean momentum equation to obtain \eqref{aa} and hence \eqref{wav3}. Restoring this term leads to the generalisation
\beq
(\partial_t + \lie_{\buL}) \left( \buE \bcdot \dd \bx + \tfrac{1}{2}  \dd ( \p_t \overline{|\bq_1|^2} ) + \tfrac{1}{2} (\bm{f} \times \bx)   \bcdot \dd \bx \right) = - \dd \piL \com \label{muone}
\eeq
of the CL equation valid for $\mu=O(1)$. Since $\lie_{\buL}$ and $\dd$ commute, $\lie_{\buL} \dd ( \p_t \overline{|\bq_1|^2} ) = \dd (\cdots)$ and the additional terms involving $|\bq_1|^2$ can be absorbed in the differential of the pressure-like term on the right-hand side. We conclude that the CL equation \eqref{wav3} with $\buSsol$ as Stokes velocity holds for $\mu=O(1)$ provided that the effective pressure $\varpi$ is suitably redefined.

%\bigskip
%
%\textcolor{red}{Can we  present  the expression for $\varpi$ in \eqref{wav3}? Leibovich (1980) gives an expression which does not seem to be the same as Suzuki \& Fox-Kemper. In a free-surface flow there are energy transfer associated with $\varpi$ so we should not dismiss the form of the modified pressure as inconsequential. According to  Suzuki \& Fox-Kemper
%$$
%\varpi = p + \half |\buL|^2 - \half |\buE|^2\, .
%$$
%This expression for $\varpi$ makes the Lagrangian energy equation in \eqref{vent17} below very pretty. Can we obtain this from \glm?}
%\JVcom{This can be done. But I'd rather not do it for this paper.}
%\textcolor{red}{It is  easier to do this now, rather than later. Below \eqref{Eulerform} we have
%$$
%\pi = p - \tfrac{1}{2} |\bu|^2 - \tfrac{1}{2}(\bm{f} \times \bx) \bcdot \bu\per
%$$
%With the $- \tfrac{1}{2} |\bu|^2$ it seems that we are already half way to Suzuki \& Fox-Kemper's expression. Is the algebra arduous?}

\section{Conclusion} \label{sec:conclusion}

This paper examines a problematic aspect of the Stokes velocity $\buS$, namely its divergence $\div \buS$ and non-zero vertical component $\wS$. Some confusion has arisen because $\div \buS$ and $\wS$ are small when the wave field has an amplitude that varies on scales longer and slower  than  the wavelength and period. This scale-separation  approximation corresponds  to the existence of a small parameter $\mu \ll 1$, as is the case for slowly varying wavepackets. The distinction between approximate and exact results in the literature  is not always made plain. Beyond this, there are no irretrievable difficulties with the familiar form \eqref{eq2} of $\buS$: when $\div \buS$ and $\wS$ cannot be neglected, they are readily computed in terms of  first-order fields.

The main point of the paper, however, is that the Stokes velocity, understood as the difference $\buS = \buL - \buE$ between Lagrangian- and Eulerian-mean velocities, is not uniquely defined and that an alternative version, $\buSsol$ in \eqref{uSsol}, that is exactly solenoidal can serve as a convenient substitute for the familiar \eqref{eq2}. The non-uniqueness arises because there is no single definition of the Lagrangian-mean velocity $\buL$, which is only constrained to serve as a good `guiding-centre' representative of the motion of an ensemble of fluid particles. The ambiguity is usually resolved by imposing the GLM condition $\overline{\bxi}=0$, which is restricted to Euclidean geometry or depends on coordinates,  but alternatives which have the benefit of both coordinate-independence and non-divergent $\buL$ exist \cite{gilb-v18}. One of these, Soward \& Roberts' glm \cite{sowa-robe10}, leads to $\buSsol$ in \eqref{uSsol} or, in a more geometric form, \eqref{eq:buSsolgeo}. We give only the form of $\buSsol$ to leading order in the wave amplitude parameter $\ep$; as we describe, higher-order corrections can be computed systematically using a Lie series expansion. We further show that the Craik--Leibovich (CL) equations \cite{crai-leib76,leib80}, widely used to model the feedback of surface gravity waves on the flow, apply in the glm framework, with $\buSsol$ simply replacing $\buS$.

We stress that, in general, the coordinates associated with a Lagrangian-mean trajectory $\bxL = \bphiL(\bx_0,t,\ep)$ of given fluid particle (identified by  initial position $\bx_0$) are not the average of the coordinates of this particle. This is a feature specific to GLM, intimately connected to its coordinate dependence and the decision to make $\overline{\bxi}=0$.

We conclude by noting that the assumption $\mu \ll 1$ which makes it possible to ignore the divergence and vertical component of $\buS$, is restrictive. The assumption clearly holds for single wavepackets, but may fail for more complex wave fields. When it does, using the solenoidal Stokes velocity $\buSsol$ brings considerable simplifications, e.g.\ for the computation of the Eulerian mean flow from the CL or other wave-averaged equations.

%\ethics{JV and WRY are ethical chaps}
\medskip

\noindent
\textbf{Acknowledgments.}{JV is supported by the UK Natural Environment Research Council, grant NE/W002876/1. WRY is supported by National Science Foundation Award OCE-2048583. We thank  Oliver B\"uhler and an anonymous referee for comments that improved this work.}

%\ack{We thank some useless buggers.}

\bibliographystyle{plain}
%\bibliography{StokesDriftDiscontents}

\appendix

\section{Computational details for \S\ref{sec:glm}} \label{app:details}

\subsection{Derivation of \eqref{dw}}

We define $\hat \bq = \bXi^*\bq $ in analogy with $\hat \bw = \bXi^* \bw$ and note that both are the negative of velocities associated with the inverse map $\bXi^{-1}$ in the sense that
\beq
\partial_t \bXi^{-1}(\bx,t,\ep) = - \hat \bw(\bXi(\bx,t,\ep),t,\ep) \quad \textrm{and} \quad \partial_\ep \bXi^{-1}(\bx,t,\ep) = - \hat \bq(\bXi(\bx,t,\ep),t,\ep),
\label{eq:hatwhatq}
\eeq 
as follows from the differentiation of the identity $\bXi^{-1} (\bXi(\bx,t,\ep),t,\ep) = \bx$ with respect to $t$ and $\ep$. (We do not make the dependence on $\alpha$ explicit for simplicity.)
Equating the derivative of the first equality with respect to $\ep$ with the derivative of the second with respect to $t$ gives
\beq
- \partial_\ep \hat \bw + \hat \bq \bcdot \grad \hat \bw = - \partial_t \hat \bq + \hat \bw \bcdot \grad \hat \bq
\eeq
and, on rearranging, 
\beq
\partial_\ep \hat \bw  = \partial_t \hat \bq + \hat \bq \bcdot \grad \hat \bw  - \hat \bw \bcdot \grad \bq = \partial_t \hat \bq + \lie_{\hat \bq} \hat \bw,
\eeq
i.e.\ \eqref{dw}.

\subsection{Equivalence between  \eqref{Euler} and \eqref{Eulerform}}

We show that \eqref{Eulerform} is equivalent to the familiar form \eqref{Eulerform} of the Euler equation. We first write \eqref{Eulerform} explicitly as
\beq
(\partial_t + \lie_{\bu}) \left( \bu \bcdot \dd \bx + \tfrac{1}{2} ( \bm{f} \times \bx) \bcdot  \dd \bx \right)= -\dd \left( p - \tfrac{1}{2} |\bu|^2 -  \tfrac{1}{2} ( \bm{f} \times \bx) \bcdot \bu \right).
\label{Eulerunpack}
\eeq
Expanding, we find
\begin{align}
& (D_t \bu) \bcdot \dd \bx  + \bu \bcdot \dd \bu +  \tfrac{1}{2} (\bm{f} \times \bu) \bcdot \dd \bx + \tfrac{1}{2} (\bm{f} \times \bx) \bcdot \dd \bu \nonumber \\
&= - \nabla p \bcdot \dd \bx + \bu \bcdot \dd \bu + \tfrac{1}{2} (\bm{f} \times \dd \bx) \bcdot \bu +  \tfrac{1}{2} (\bm{f} \times \bx) \bcdot \dd \bu,
\end{align}
which recovers \eqref{Euler} since $(\bm{f} \times \dd \bx) \bcdot \bu = - (\bm{f} \times \bu) \bcdot \dd \bx$. 

%with $D_t = \partial_t + \bu \cdot \nabla$, hence the familiar
%\beq
%D_t \bu+  2 \bm{\Omega} \times \bu = - \nabla p.
%\eeq
%Eq.\ \eqn{Eulerunpack} gives the coordinates expressions

\subsection{GLM from \eqref{avEuler}}

We show that \eqref{avEuler}, together with the defining property of GLM $\bar \bxi = 0$ is equivalent to Theorem I of Andrews \& McIntyre \cite{andr-mcin78a} in the case of an incompressible fluid. We need to apply the pull-back $\bXi^*$ to 
\beq
\nu_a = \bu \bcdot \dd \bx + \tfrac{1}{2} (\bm{f} \times \bx) \bcdot  \dd \bx  \quad \textrm{and} \quad
\pi = p - \tfrac{1}{2} |\bu|^2 - \tfrac{1}{2} (\bm{f} \times \bx) \bcdot \bu,
%\label{nuapi}
\eeq
then average.
Here $\bu$ stands for the 3 components of $\bu$ rather than for a vector, hence the pull-back is simply composition with $\bXi$, that is, $\bXi^* \bu = \bu \circ \bXi =: \bu^\xi$. We also have $\bXi^* \bx = \bx + \bm{\xi}(\bx,t,\ep)$ and $\bXi^*  \dd \bx = \dd (\bXi^* \bx) = \dd \bx + \dd \bxi$. Using that $\bar \bxi = 0$, we compute
\beq
\nu_a^\mathrm{L} = \overline{\bu^\xi \bcdot \dd \bx + \bu^\xi \bcdot \dd \bxi + \tfrac{1}{2} (\bm{f} \times \bxi) \bcdot  \dd \bxi} + \tfrac{1}{2} (\bm{f} \times \bx) \bcdot  \dd \bx.
\label{eq:nua}
\eeq 
The averaged term can be written as $(\buL - \bm{\mathsf{p}}) \bcdot \dd \bx$ with $\buL = \overline{\bu^\xi}$ and
\beq
\mathsf{p}_i = - \overline{ \xi_{j,i} (u^\xi_j   +\tfrac{1}{2} \left(\bm{f} \times \bxi)_{j}\right)}
\eeq
the pseudomomentum.
Similarly,
\beq
\piL = \overline{{p}^\xi - \tfrac{1}{2} |\bu^\xi|^2 - \tfrac{1}{2} \left(\bm{f} \times \bxi \right) \bcdot \bu^\xi} - \tfrac{1}{2}\left(\bm{f} \times \bx\right) \bcdot \buL.
\label{eq:pia}
\eeq
Introducing \eqref{eq:nua} and \eqref{eq:pia} into \eqref{avEuler}, we compute
\begin{align}
(\partial_t + \lie_{\buL}) \left((\buL - \bm{\mathsf{p}}) \bcdot \, \dd \bx \right) & + \left( \left( \curl \tfrac{1}{2} (\bm{f} \times \bx) \right) \times \buL \right)  \bcdot  \dd \bx  + \dd \left( \tfrac{1}{2} (\bm{f} \times \bx) \bcdot \buL \right) \nonumber \\ & = - \nabla \pitL \bcdot \dd \bx  + \dd \left( \tfrac{1}{2} (\bm{f} \times \bx) \bcdot \buL \right),
\end{align}
where $\pitL = \overline{{p}^\xi - \tfrac{1}{2} |\bu^\xi|^2 - \tfrac{1}{2} (\bm{f} \times \bxi) \cdot \bu^\xi}$ and we have used Cartan's formula given below \eqref{aa}. Since $\curl (\bm{f} \times \bx) = 2 \bm{f}$  we finally obtain
\beq
D_t (u^\mathrm{L}_i - \mathsf{p}_i) + (u^\mathrm{L}_j - \mathsf{p}_j) u^\mathrm{L}_{j,i} +  (\bm{f} \times \bu^\mathrm{L})_i = - \tilde \pi^\mathrm{L}_{,i},
\eeq
that is, Theorem I of \cite{andr-mcin78a}.

\end{document}